\documentclass[a4paper,12pt]{article}

\usepackage[a4paper,margin=1in]{geometry}
\usepackage{multirow}
\usepackage{longtable}
\usepackage{array}
\usepackage{times}
\usepackage{setspace}
\usepackage{graphicx}
\usepackage{amsmath,amssymb}
\usepackage{booktabs}
\usepackage{url}
\usepackage{hyperref}
\usepackage{caption}
\usepackage{float}
\usepackage{tabularx}
\usepackage{booktabs}
\usepackage{makecell}
\usepackage{array}
\newcolumntype{C}{>{\centering\arraybackslash}X}
\begin{document}
\title{A Multi-Scenario UAV RF Dataset with Real-World Acquisition and Signal Processing Benchmarking}

\author{Haolin Zheng$^{1}$, Ning Gao$^{*1}$, Zhenghang Zhu$^{1}$, Zhijun Huang$^{2}$,\\ Shi Jin$^{3}$, Michail Matthaiou$^{4}$}

\date{
	$^1$School of Cyber Science and Engineering, Southeast University, Nanjing 210096, China\\
	$^2$Chien-Shiung Wu College, Southeast University, Nanjing 210096, China\\
	$^3$National Mobile Communications Research Laboratory, Southeast University, Nanjing 210096, China\\
	$^4$Centre for Wireless Innovation (CWI), Queen's University Belfast, Belfast BT3 9DT, U.K.\\
    $^*$Corresponding Author. Email:~ninggao@seu.edu.cn
}

\maketitle

\begin{abstract}
We present a real-world multi-scenario unmanned aerial vehicle (UAV) radio frequency (RF) dataset, namely DRFF-R2, which is collected using a dedicated acquisition platform under diverse operational conditions. All signals are acquired within a unified framework to ensure consistency in hardware configuration and environmental settings. The dataset is systematically organized into seven well-defined subsets corresponding to different operational and signal composition scenarios to facilitate structured experimentation. Each file follows a clearly annotated naming convention to enable convenient data indexing and reproducible analysis. The dataset contains RF recordings from 26 UAV units spanning 8 distinct models, captured across varying flight states, altitudes, speeds, acquisition days, and receiver configurations. By covering diverse acquisition settings and signal compositions, the dataset provides a comprehensive resource for future UAV RF signal research, including RF fingerprinting (RFF) identification, model-level recognition, flight state analysis, time-varying RFF study, and interference-aware signal processing.
\end{abstract}

\section*{Background \& Summary}

The widespread adoption of low-altitude unmanned aerial vehicles (UAVs, also known as drones) across commercial, industrial, and civilian sectors has brought significant benefits to modern society. However, this rapid growth in drone deployment has also led to an increase in unauthorized and illegal activities, such as restricted airspace intrusion, illegal transportation, and privacy violations~[1]. To address these concerns and ensure safe drone regulation, one promising approach is radio frequency fingerprinting (RFF), which leverages inherent hardware imperfections introduced during the manufacturing process~[2]. Despite its potential, the study of drone RFF relies heavily on the extraction of unique features from drone samples, which in turn depends on the availability of large and diverse datasets. Unfortunately, the lack of a large-scale drone radio frequency (RF) dataset presents a critical gap in this field.

Several research teams have responded to this challenge by publishing and open-sourcing their independently collected drone RF datasets. One representative dataset is DroneRFa~[3], which includes RF signals from 9 types of flying drones recorded in an outdoor environment, 15 types recorded indoors, and 1 type of background signal for reference. Each dataset contains at least 12 segments, with more than 100 million sampling points per segment. Another dataset, DroneRFb~[4], extended DroneRFa by focusing on non-cooperative drone control, with RF signal data from 6 different types of drones in an urban environment, as well as a reference background signal. This dataset was collected using software-defined radio (SDR) equipment and includes detailed annotations for individual identifiers, along with line-of-sight (LoS) or non-line-of-sight (NLoS) conditions. The DRFF-R1~[5] dataset comprises RF data collected from 20 UAVs across 7 different model types. The dataset covers various flight conditions, including different altitudes and velocities, and was acquired in real electromagnetic environments. It captures practical channel effects, including Doppler shifts, reflecting realistic signal propagation characteristics. The RFUAV dataset~[6] offers approximately 1.3 TB of raw frequency data from 37 distinct drones, providing a comprehensive benchmark for drone detection and identification. It addresses many limitations of previous datasets by including a wide variety of drone types, large data volumes, varying signal-to-noise ratios (SNRs), and open-access evaluation tools. Additionally, the Vuorenmaa \textit{et al.}~[7] collects I/Q-baseband recordings of radio control and video signals from 10 consumer/prosumer drones, captured at center frequencies of 2.44 GHz and 5.8 GHz. These recordings are made in the anechoic chamber, with drones placed on a turntable to ensure consistent signal capture. And, this dataset also provides the complete hopping sequence of the control signal, and in some cases, multiple recordings for a single drone. These datasets represent important resources in advancing drone RFF research, enabling the studies of drone detection, classification, and tracking systems.

However, research on UAV RF signals has progressed beyond the classification of single drones to encompass more advanced tasks, including UAV flight state awareness~[8][9][10], drone recognition under interference  conditions~[11][12][13][14], and the analysis of time-varying RFF characteristics~[15][16]. These emerging research directions place new demands on dataset design, requiring unified data acquisition under diverse operational scenarios. Although the existing datasets provide substantial data for specific scenarios, they lack widespread coverage of diverse drone operation scenarios and systematic investigations into the various factors influencing UAV RFF characteristics. A detailed comparison between these datasets and DRFF-R2 is presented in Table~1. Moreover, these requirements cannot be adequately addressed by simply aggregating existing datasets, as variations in acquisition hardware, UAV models, environmental conditions, and signal configurations introduce inconsistencies that undermine cross-task comparability and unified evaluation.

To address these challenges, we hereafter introduce DRFF-R2, a large-scale UAV RF dataset collected under diverse operational, propagation, and acquisition conditions to support a broad range of RF sensing and learning tasks. Specifically, for deep-learning-based RFF identification, such as UAV individual identification and type recognition, DRFF-R2 provides extensive I/Q recordings from multiple UAV devices, models, operating states, and scenarios, enabling the learning of discriminative device-dependent representations~[17][18]. The dataset also includes both single-UAV recordings and challenging mixtures, such as simultaneous multi-UAV emissions and WiFi-UAV coexistence signals, providing paired source and mixture observations for developing and quantitatively evaluating learning-based RF source-separation methods inspired by architectures such as KUIELab-MDX-Net~[19], Hybrid Transformer Demucs~[20], and Wave-U-Net~[21]. Moreover, given that RFF representations can be sensitive to receiver hardware, wireless channels, and acquisition environments~[22][23][24], DRFF-R2 explicitly incorporates naturally occurring domain variations in acquisition time, UAV height, flight speed, receiver configuration, and indoor/outdoor environments, supporting both domain-adaptation and domain-generalization studies. Thus, the multiple scenarios in DRFF-R2 are designed not merely to increase data volume, but also to expose RFF algorithms to diverse sources of intra-device and inter-domain variation, enabling systematic investigation of whether learned representations capture persistent device-specific characteristics or condition-dependent features.

\begin{table*}[t]
	\footnotesize
	\centering
	
	\caption{Comparison of representative UAV RF signal datasets.}
	\label{tab:dataset_comparison}
	
	\renewcommand{\arraystretch}{1.25}
	\setlength{\tabcolsep}{3.5pt}
	
	\begin{tabular}{
			>{\centering\arraybackslash}m{2.3cm}
			>{\centering\arraybackslash}m{1.4cm}
			>{\centering\arraybackslash}m{1.6cm}
			>{\centering\arraybackslash}m{4.0cm}
			>{\centering\arraybackslash}m{1.6cm}
			>{\centering\arraybackslash}m{3.9cm}
		}
		
		\toprule
		
		Dataset &
		Dataset size &
		Number of UAVs &
		Covered scenarios &
		Frequency bands &
		Applicable research
		\\
		
		\midrule
		
		DroneRFa~[3] &
		$\sim$574 GB &
		16 UAVs / 16 types &
		Indoor/outdoor environments with different altitudes and velocities &
		915 MHz, 2.4 GHz, 5.8 GHz &
		Type classification
		\\
		
		\addlinespace[6pt]
		
		DroneRFb~[4] &
		$\sim$164 GB &
		18 UAVs / 6 types &
		LoS/NLoS urban environments with interference &
		2.4 GHz &
		Individual identification, type classification
		\\
		
		\addlinespace[6pt]
		
		DRFF-R1~[5] &
		$\sim$136 GB &
		20 UAVs / 7 models &
		Outdoor environments with different altitudes &
		5.8 GHz &
		Individual identification, type classification
		\\
		
		\addlinespace[6pt]
		
		RFUAV~[6] &
		$\sim$1.3 TB &
		37 UAVs / 37 types &
		Real-world scenarios with various SNR conditions &
		2.4 GHz, 5.8 GHz &
		\makecell[c]{
			Type classification, \\
			SNR robustness evaluation
		}
		\\
		
		\addlinespace[6pt]
		
		Vuorenmaa \textit{et al.}~[7] &
		$\sim$9 GB &
		10 UAVs / 10 types &
		Dual-band frequency-hopping signals acquired in controlled environments &
		2.4 GHz, 5.8 GHz &
		\makecell[c]{
			Type classification,\\
			frequency-hopping analysis
		}
		
		\\
		
		\addlinespace[6pt]
		
		\textbf{DRFF-R2 (ours)} &
		$\sim$375 GB &
		26 UAVs / 8 models &
		Multi-UAV co-channel, multiple flight states, indoor/outdoor environments with altitude and velocity variations, cross-day and cross-receiver measurements, dual-band frequency hopping, Wi-Fi interference &
		2.4 GHz, 5.8 GHz &
		\makecell[c]{
			Type classification, \\
			individual identification, \\
			flight-state recognition, \\
			interference-aware analysis, \\
			time-varying RFF modeling, \\
			cross-domain evaluation, \\
			frequency-hopping analysis, \\
			co-channel signal separation
		}
		
		\\
		
		\bottomrule
		
	\end{tabular}
	
\end{table*}

\section*{Methods}

The dataset was constructed through a controlled RF reception procedure comprising environmental baseline measurement, scenario-specific data collection and structured data storage. Data are shared under a CC-BY 4.0 license in a public repository (see Data Records section for details).

\subsection*{Equipment Preparation and System Configuration}

The hardware components were configured and tested prior to data collection. The acquisition system consisted of the UAVs under test, RF receivers, a data storage subsystem, and auxiliary field equipment. As shown in Figure~1, all components were installed and operated according to predefined configuration settings.

\begin{figure}[h]
	\centering
	\includegraphics[width=16cm]{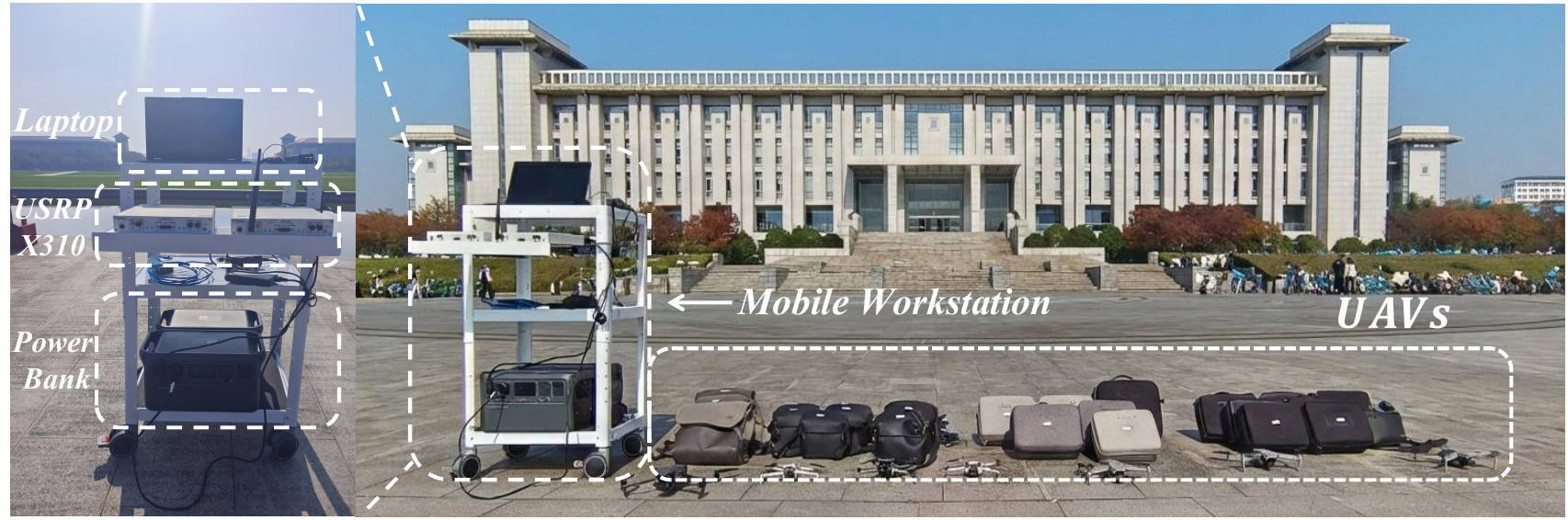}
	\caption{Overall hardware architecture of the RF signal acquisition system, including UAVs, SDR-based receiver, data storage device, and  auxiliary equipment.}
	\label{fig:system_architecture}
\end{figure}

\begin{itemize}
	
	\item \textbf{UAVs:}
	The dataset comprises eight mainstream consumer-grade DJI drone models, totaling 26 individual airframes, as summarized in Table~1. For several models, multiple units were considered to capture intra-model variability and enhance the diversity of device-level RF fingerprint characteristics within the dataset.
	
	\begin{table}[h]
		\centering
		\caption{Drone platforms included in the dataset}
		\label{uav}
		\begin{tabular}{cccc}
			\hline
			Manufacturer & Model & Quantity \\
			\hline
			DJI & Mini 5 Pro & 1  \\
			DJI & Mini 4 Pro & 5 \\
			DJI & Mini 3 Pro & 2 \\
			DJI & Mavic 3 & 1 \\
			DJI & Mavic 3C & 1 \\
			DJI & Mavic Air 3S & 1 \\
			DJI & Mavic Air 2S & 7 \\
			DJI & Mavic Air 2 & 8 \\
			\hline
			Total & 8 Models & 26 Units &  \\
			\hline
		\end{tabular}
	\end{table}
	
	\item \textbf{RF Receiver:}
	RF signal acquisition was performed using SDR receivers consisting of two Ettus Research universal software radio peripheral (USRP) 2943 devices. The receivers support wide instantaneous bandwidth and high-rate streaming. Prior to each acquisition session, device connectivity was checked, clock synchronization was configured, and acquisition parameters were confirmed to match predefined settings. Baseband signals were recorded under fixed hardware and configuration conditions.
	
	\item \textbf{Data Storage Subsystem:}
	The SDR receivers were connected to a Windows 11 laptop equipped with 16GB RAM and a 2TB external storage device. Prior to each acquisition session, buffer allocation and disk write performance were tested under the configured sampling conditions. System resource utilization was monitored during recording.
	
	\item \textbf{Auxiliary Filed Equipment:}
	A dual-band omnidirectional antenna with a gain of 5~dBi was used for RF signal reception. Antenna orientation, mounting height, and spatial position relative to the UAV flight path were kept constant across acquisition sessions.
	
	For outdoor experiments, the acquisition system was deployed on a mobile workstation platform and powered by portable battery units. The power supply configuration was maintained consistently during field measurements.
	
	\item \textbf{Acquisition Software and Streaming Implementation:}
	Data acquisition was controlled using Python scripts developed on top of the UHD driver framework. All signal labels and associated metadata fields were predefined within the acquisition scripts and loaded automatically during recording (see Data Records section for details). Label information was written to file headers together with the corresponding signal segments. The acquisition program is publicly available at \url{https://doi.org/10.57760/sciencedb.36815}.
	
	During signal acquisition, the sampling rate was fixed at \( F_s = 100\,\mathrm{MS/s} \). Each acquisition segment contains approximately
	\(1.4 \times 10^{8}\) complex I/Q samples. Moreover, the center frequency was configured according to the experimental scenario. For outdoor signal acquisition, a center frequency of 5.745\,GHz was used. For indoor  WiFi-interference signal acquisition, the center frequency was set to 2.437\,GHz.
	
\end{itemize}

\subsection*{Scenario Design}

Data was collected across multiple operational scenarios to capture realistic propagation dynamics and interference conditions:

\begin{enumerate}
	\item \textbf{Ambient Noise Scenario:}
	Prior to UAV activation, background RF signals were recorded to characterize the baseline electromagnetic environment. These recordings serve as reference data for spectrum occupancy analysis, noise power estimation, and SNR assessment.
	
	\item \textbf{Controlled Propagation Scenario:}
	Signal collection was conducted inside an enclosure lined with RF absorbing material. This configuration was used to constrain multipath propagation and external interference, thereby emulating a quasi free-space transmission condition.
	The recorded samples correspond to RF emissions captured under reduced environmental channel variability.
	
	\item \textbf{Flight Scenarios:}
	Controlled flight maneuvers were performed at predefined distances (e.g., 50\,m horizontal separation from the receiver) and altitudes (10\,m, 30\,m, and 50\,m). The operational states included hovering, straight-line motion, curved trajectories, takeoff, and landing. These configurations introduce realistic flight-state representations, including time-varying path loss, Doppler shifts due to platform motion, and control related spectrum changes.
	
	\item \textbf{Interference Scenarios:}
	Signal collection was conducted in enclosed indoor environments with active WiFi traffic in the 2.4\,GHz band. The UAV operated within the same frequency range, resulting in co-channel and adjacent-channel interference conditions. The recorded samples therefore reflect RF emissions captured under spectrally congested WiFi environments.
	
	\item \textbf{Multi-Drone Mixed Scenarios:}
	Multiple UAVs were utilized simultaneously within the same frequency band during data collection. Concurrent transmissions resulted in partial or complete spectral overlap, time-varying power levels, and composite signal superposition at the receiver. The recorded samples therefore contain mixed RF emissions from multiple transmitters captured under the same frequency spectrum.
	
\end{enumerate}

These scenarios were predefined during the acquisition process to enable task-oriented dataset organization, allowing different subsets of the collected data to be tailored for specific scientific questions and downstream UAV RF sensing tasks (see the Data Records section for details).

\subsection*{Signal Acquisition Procedure}

As illustrated in Figure~2, all data collection sessions followed a predefined acquisition workflow covering UAV operation, SDR configuration, and data recording.

The UAV operators were responsible for drone operation and flight-state control. Prior to recording, the UAV was powered in proximity to the receiving antenna, and a communication link was established via a calibrated remote controller or the official mobile application. Once connected, telemetry status, GPS positioning, video transmission quality, and battery level were directly monitored through the application interface to confirm stable operating conditions. The operators executed predefined UAV flight states, including hovering, linear motion, curved trajectories, takeoff, and landing, at specified distances and altitudes according to the scenario configuration.

The acquisition program operator was responsible for SDR configuration and data collecting. The acquisition workstation was initialized, and the USRP device was connected and configured using a UHD-based Python program. Key parameters, including sampling rate, center frequency, and receiver gain, were configured according to the designated scenario and fixed prior to recording. During acquisition, signal quality was monitored in real time through time-frequency-domain observations. Specifically, qualitative indicators, including time-frequency intensity distributions, signal boundary stability, and repetitive signal structures, were examined to identify abnormal interference or acquisition instability. Since the collected signals were intended for passive RF fingerprinting analysis, no protocol-dependent demodulation procedures, such as constellation analysis or bit-error-rate evaluation, were performed. Upon completion of the current acquisition procedure, the recorded data file was saved using a structured naming convention embedding essential metadata, after which the workflow proceeded to the next predefined acquisition procedure.

It should be emphasized that the collected data contains the original complex I/Q recordings directly captured from the SDR receiver. No additional signal processing operations, including synchronization, burst detection, packet extraction, carrier frequency offset (CFO) compensation, IQ imbalance correction, DC offset removal, normalization, filtering, or segmentation, were applied during the data acquisition and recording process. The dataset aims to preserve the original RF characteristics and avoid potential modification or removal of device-dependent RFF information caused by preprocessing operations. For downstream applications, researchers can perform task-specific preprocessing operations, such as filtering, normalization, downsampling, and segmentation, according to their specific algorithms, model architectures, and evaluation requirements.

\begin{figure}[h]
	\centering
	\includegraphics[width=16cm]{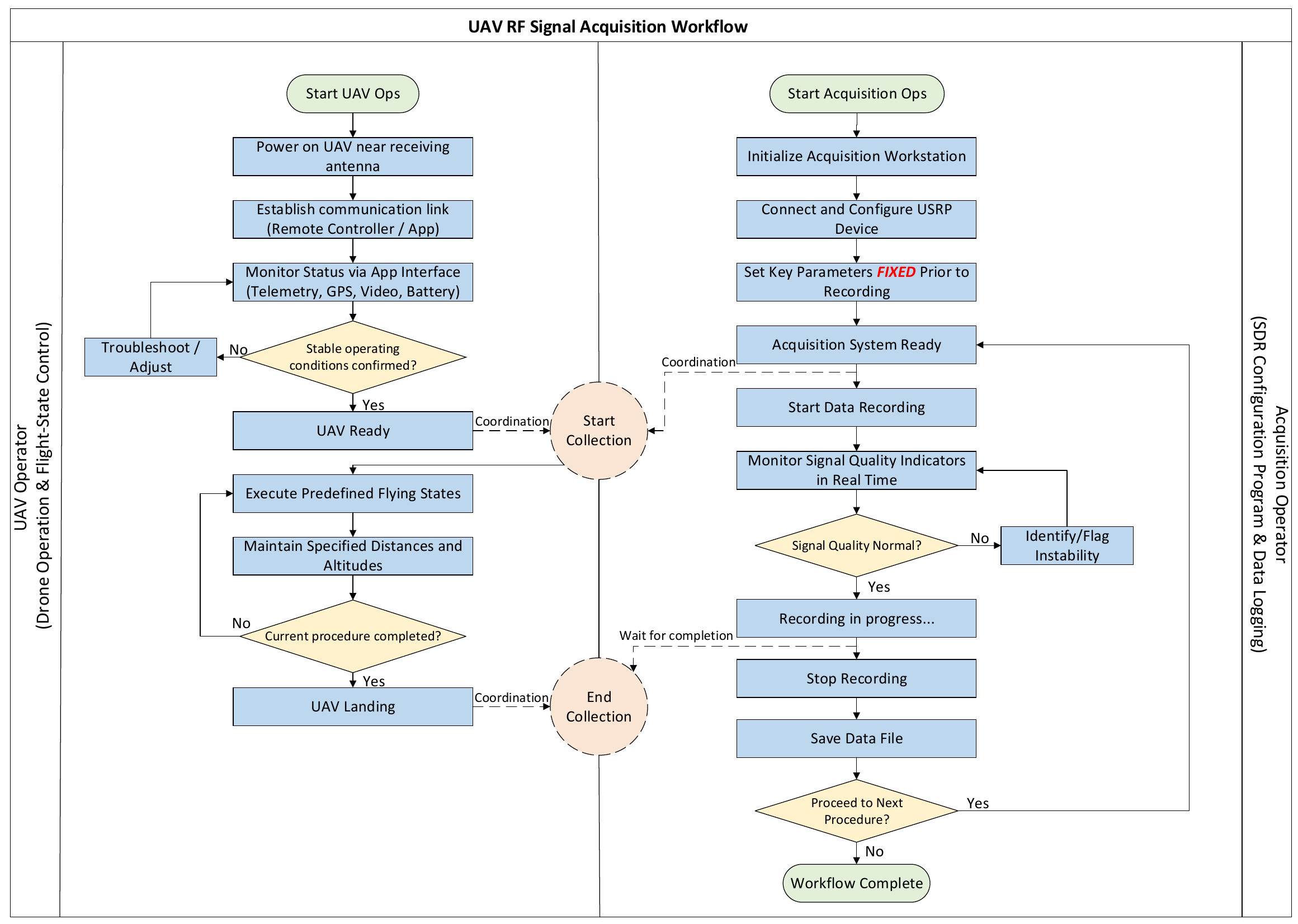}
	\caption{Standardized signal acquisition flowchart.}
	\label{fig:acquisition_workflow}
\end{figure}
\section*{Data Records}
The dataset is publicly hosted on the Science Data Bank platform at \url{https://doi.org/10.57760/sciencedb.36815}~[25]. This dataset is distributed under the Creative Commons Attribution 4.0 International (CC BY 4.0) license (\url{https://creativecommons.org/licenses/by/4.0/legalcode}).

It consists of seven directories, each designed to address specific scientific questions in UAV RF sensing by isolating key factors that affect RFFs, including UAV behavioral variations, multi-source interference, channel dynamics, frequency hopping, and environmental effects.

\begin{itemize}
	\item \texttt{dataset1-single\_drone\_states}: This directory contains RF signal data collected from a single UAV under different operational states, including cruise, takeoff, ascend, descend, landing, and shading. This dataset enables research on UAV flight-state recognition based on passive RF observation, which can support practical applications such as detecting abnormal flight behaviors, unauthorized aerial filming, or illegal landing activities.
	
	\item \texttt{dataset2-drone\_mixed}: This directory contains RF signals from multiple UAVs transmitting simultaneously. Each subdirectory corresponds to a specific multi-UAV combination. For example, \texttt{mavic\_air2s\_2\_\&\_mavic\_air2s\_3} contains mixed RF signals collected from two individual Mavic Air 2S UAVs. This dataset supports the development of robust UAV detection algorithms under strong co-channel interference conditions and provides a benchmark for investigating multi-UAV signal separation and interference mitigation techniques.
	
	\item \texttt{dataset3-single\_drone\_hover}: This directory provides UAV RF signals collected during hovering operations across different days (e.g., Day 1, Day 2, and Day 3), receiver devices, altitudes, and velocities. These variations introduce realistic temporal and hardware-domain shifts, enabling research on model generalization, cross-domain adaptation, and robust RF fingerprint modeling.
	
	\item \texttt{dataset4-single\_drone\_dual\_frequency}: This directory contains RF signal data collected when UAVs operate in frequency-hopping modes. The RF signals outside this directory are collected without frequency hopping enabled. This dataset facilitates research on UAV detection and RF fingerprint extraction under dynamic frequency variations.
	
	\item \texttt{dataset5-single\_drone\_inside\_absorbent\_cotton}: This directory contains RF signals collected from a UAV enclosed by electromagnetic absorbent cotton. The absorbent material reduces external electromagnetic reflections and suppresses environmental interference, thereby mitigating propagation effects such as, multipath reflections, NLoS components, and background noise. Moreover, the close placement between the UAV and the receiver reduces propagation path variations and provides relatively isolated UAV RF observations. These measurements are valuable for investigating intrinsic UAV RF characteristics, hardware-related fingerprint components, and RFF modeling. The directory further includes \texttt{dual\_frequency\_5.745\&2.437} and \texttt{single\_frequency\_5.745}, representing signals collected with and without frequency hopping, respectively.
	
	\item \texttt{dataset6-wifi\_mixed}: This directory contains RF signals collected from UAVs under Wi-Fi interference conditions. It enables research on UAV identification and RF fingerprint extraction in the presence of strong external electromagnetic interference.
	
	\item \texttt{dataset7-environment}: This directory contains environmental RF signals collected before UAV signal acquisition. These background measurements characterize external electromagnetic conditions and can be used for background channel subtraction and environmental bias reduction to improve RF signal quality.
\end{itemize}

The drone RF signal data are stored in MAT format. To better organize and store the UAV RF signal data, we have clearly annotated the naming format for each file. Specifically, consider the file name \texttt{mavic3\_1\_Ascend\_c17\_u1\_d2.mat}:
\begin{itemize}
	\item \texttt{mavic3\_1}: Indicates the first UAV instance of the MAVIC 3 model.
	\item \texttt{\_Ascend}: Represents the current flight state of the drone.
	\item \texttt{\_c*}: Refers to a combination of the drone's flight mode, speed, and state. Since the combinations vary across different dataset dictionaries, we have provided detailed information about these combinations in Table~2 for reference.
	\item \texttt{\_u*}: Denotes the USRP number used for collecting the signal.
	\item \texttt{\_d*}: Indicates the day on which the signal was captured.
\end{itemize}
\begin{longtable}{p{1.5cm} c c c c c c}
	\caption{Description of UAV RF signal datasets and flight parameters} \label{tab:drone_datasets} \\
	\toprule
	\textbf{Dataset} & \textbf{Label} & \textbf{Alt. (m)} & \textbf{Speed (m/s)} & \textbf{Flight Maneuver} & \textbf{Lat. Dist. (m)} & \textbf{State} \\
	\midrule
	\endfirsthead
	
	\multicolumn{7}{c}%
	{{\bfseries \tablename\ \thetable{} - continued from previous page}} \\
	\toprule
	\textbf{Dataset} & \textbf{Label} & \textbf{Alt. (m)} & \textbf{Speed (m/s)} & \textbf{Flight Maneuver} & \textbf{Lat. Dist. (m)} & \textbf{State} \\
	\midrule
	\endhead
	
	\midrule
	\multicolumn{7}{r}{{Continued on next page}} \\
	\bottomrule
	\endfoot
	
	\bottomrule
	\multicolumn{7}{l}{\footnotesize \textit{Note: Lat. Dist. = Lateral Distance to Receiver; Alt. = Altitude.}}
	\endlastfoot
	
	\multirow{26}{*}{\parbox{3cm}{\textbf{Dataset 1}}}
	& c1 & 10 & 0-5 & Linear Flight & 50 & Cruise \\
	& c2 & 10 & 5-10 & Linear Flight & 50 & Cruise \\
	& c3 & 30 & 0-5 & Linear Flight & 50 & Cruise \\
	& c4 & 30 & 5-10 & Linear Flight & 50 & Cruise \\
	& c5 & 50 & 0-5 & Linear Flight & 50 & Cruise \\
	& c6 & 50 & 5-10 & Linear Flight & 50 & Cruise \\
	& c7 & 10 & 0-5 & Curved Flight & 50 & Cruise \\
	& c8 & 10 & 5-10 & Curved Flight & 50 & Cruise \\
	& c9 & 30 & 0-5 & Curved Flight & 50 & Cruise \\
	& c10 & 30 & 5-10 & Curved Flight & 50 & Cruise \\
	& c11 & 50 & 0-5 & Curved Flight & 50 & Cruise \\
	& c12 & 50 & 5-10 & Curved Flight & 50 & Cruise \\
	& c13 & 10 & 0 & Hovering (Video) & - & Cruise \\
	& c14 & 30 & 0 & Hovering (Video) & - & Cruise \\
	& c15 & 50 & 0 & Hovering (Video) & - & Cruise \\
	& c16 & $0 \to 3$ & 0-3 & Take-off & 30 & TakeOff \\
	& c17 & $3 \to 50$ & 0-5 & Ascent & 30 & Ascend \\
	& c18 & $3 \to 50$ & 5-10 & Ascent & 30 & Ascend \\
	& c19 & $50 \to 3$ & 0-5 & Descent & 30 & Descend \\
	& c20 & $50 \to 3$ & 5-10 & Descent & 30 & Descend \\
	& c21 & $3 \to 0$ & 0-3 & Landing & 30 & Landing \\
	& c22 & 10 & 0 & Hovering (Occluded) & 30 & Shading \\
	& c23 & 30 & 0 & Hovering (Occluded) & 30 & Shading \\
	& c24 & 50 & 0 & Hovering (Occluded) & 30 & Shading \\
	& c25 & Mixed & 0-5 & Linear (Occluded) & 50 & Shading \\
	& c26 & Mixed & 5-10 & Linear (Occluded) & 50 & Shading \\
	\midrule
	
	\multirow{9}{*}{\parbox{3cm}{\textbf{Dataset 2}}}
	& c1 & 10 & 0 & Hovering & 3 & Hover \\
	& c2 & 30 & 0 & Hovering & 3 & Hover \\
	& c3 & 50 & 0 & Hovering & 3 & Hover \\
	& c10 & 10 & 0-5 & Co-directional & 50 & Cruise \\
	& c11 & 10 & 5-10 & Co-directional & 50 & Cruise \\
	& c12 & 30 & 0-5 & Co-directional & 50 & Cruise \\
	& c13 & 30 & 5-10 & Co-directional & 50 & Cruise \\
	& c14 & 50 & 0-5 & Co-directional & 50 & Cruise \\
	& c15 & 50 & 5-10 & Co-directional & 50 & Cruise \\
	\midrule
	
	\multirow{3}{*}{\parbox{3cm}{\textbf{Dataset \\   3-5}}}
	& c1 & 10 & 0 & Hovering & 3 & Hover \\
	& c2 & 30 & 0 & Hovering & 3 & Hover \\
	& c3 & 50 & 0 & Hovering & 3 & Hover \\
	\midrule
	
	\multirow{4}{*}{\parbox{3cm}{\textbf{Dataset 6}}}
	& c1 & 5 & 0 & Hovering & 3 & Hover \\
	& c2 & 10 & 0 & Hovering & 3 & Hover \\
	& c4 & 5 & 0-5 & Linear Flight & 50 & Cruise \\
	& c6 & 10 & 0-5 & Linear Flight & 50 & Cruise \\
\end{longtable}
This structured naming convention enables efficient indexing, retrieval, and organization of UAV RF data. Each filename encodes key metadata, including UAV model, operational state, acquisition scenario, and timestamp, thereby ensuring clear traceability across recording sessions. In addition to filename-level metadata, each data file contains a fixed internal structure with predefined variables as summarized in Table~4.

\begin{table}[h]
	\centering
	\caption{Internal data structure of each acquisition file}
	\label{tab:data_structure}
	\begin{tabular}{c c}
		\hline
		\textbf{Variable Name} & \textbf{Description} \\
		\hline
		
		RF0\_I & In-phase (I) component of the received signal \\[0.4em]
		
		RF0\_Q & Quadrature (Q) component of the received signal \\[0.4em]
		
		Fs & Sampling rate (Hz) \\[0.4em]
		
		CenterFrequence & Center frequency (Hz) \\[0.4em]
		
		Gain & Receiver gain (dB) \\[0.4em]
		
		State & UAV operational state label (e.g., hovering) \\[0.4em]
		
		Distance & Horizontal distance between the UAV and the receiver \\[0.4em]
		
		Height & Flight altitude at the time of signal recording \\[0.4em]
		
		FlightMode & Description of the UAV flight mode \\[0.4em]
		
		\hline
	\end{tabular}
\end{table}

\section*{Technical Validation}

In this section, we conduct a set of simple yet representative characterizations to demonstrate the quality, consistency, and usability of the collected UAV RF signals.

\subsection*{Time-Frequency Representation of RF Signals}

To examine the time-frequency characteristics under different acquisition conditions, representative samples from Dataset-2, Dataset-3, and Dataset-6 were selected for analysis. Short-time Fourier transform (STFT) representations were computed for the corresponding RF signal segments.

Let $x[n]$ denote the discrete-time complex baseband I/Q signal. The STFT is defined as
\begin{equation}
	X(m) = \sum_{n=0}^{N_{\text{FFT}}-1} x[n+mH] \, w[n] \,
	e^{-j 2\pi kn / N_{\text{FFT}}},
	\label{eq:stft}
\end{equation}
where $w[n]$ is a Hann window of length $N_{\text{FFT}} = 1024$, $H = 256$ denotes the hop size, $m$ is the time-frame index, and $k$ is the frequency bin index. For each RF segment, a fixed duration of 0.1s is selected for time-frequency analysis to maintain uniform representation across all samples. The magnitude spectrogram is computed as
\begin{equation}
	S(m) = |X(m)|,
\end{equation}
and further transformed into the logarithmic domain to enhance the visibility of fine-grained spectral structures:
\begin{equation}
	S_{\text{dB}}(m) = 10 \log_{10} \left( S(m)^2 + \epsilon \right),
\end{equation}
where $\epsilon$ is a small constant introduced for numerical stability.
\begin{figure}[H]
	\centering
	\includegraphics[width=16cm]{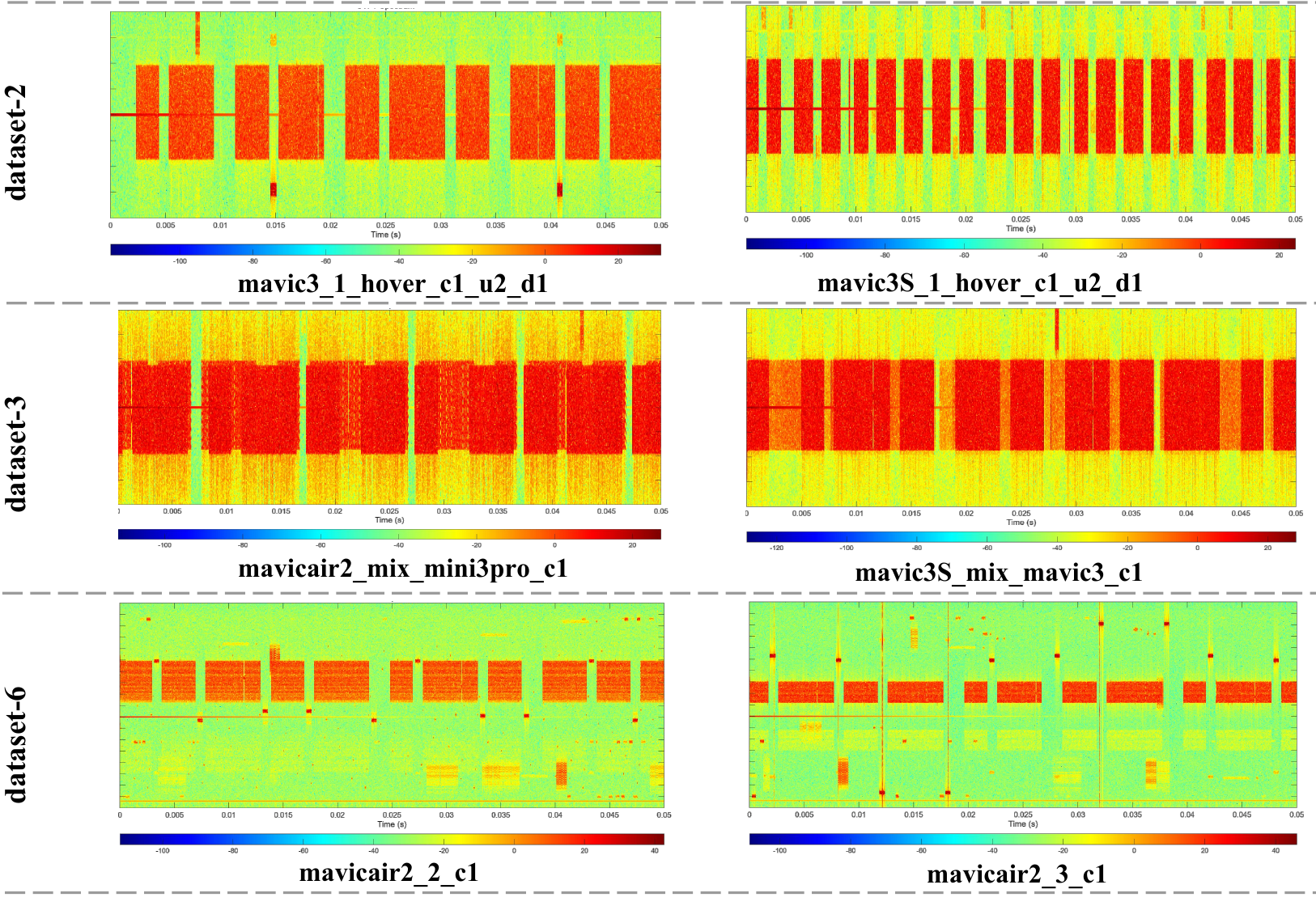}
	\caption{Representative STFT spectrogram samples from the proposed dataset under three acquisition scenarios: single-UAV hovering, dual-UAV mixed transmission, and UAV signals mixed with Wi-Fi interference.}
	\label{fig:stft_examples}
\end{figure}
Representative STFT spectrograms are shown in Figure~3. The corresponding  time-frequency representations exhibit structured spectral patterns across scenarios, with observable differences in spectral occupancy, temporal continuity, and interference structure among distinct signal compositions. No abnormal discontinuities or saturation artifacts are observed in the analyzed segments.

\subsection*{Signal Quantitative Characterization and Validation}
To quantitatively characterize and validate the received RF signals, we evaluate the recordings using five metrics: the received signal power \(P^{\mathrm{dBFS}}\), signal-to-noise ratio (SNR) \(\mathrm{SNR}^{\mathrm{dB}}\), frame-level power standard deviation \(\sigma_P\), interquartile range (IQR) \(\mathrm{IQR}_P\), and 99\% observable bandwidth \(B_{99}\). These metrics jointly characterize the received signal strength, its distinguishability from environmental noise, spectral occupancy, and short-term temporal power variation~[26]. For within-record repeatability, three non-overlapping signal segments of equal length are randomly selected from each recording. All five metrics are independently computed for each segment. The mean across the three segments is reported as the representative value for each recording, while the inter-segment standard deviation is used to quantify the consistency and repeatability of the measurements within the recording.

Specifically, the environmental recordings in Dataset-7 are used as the background-noise reference. For both the outdoor and indoor scenarios, one representative recording is randomly selected from the corresponding dataset. Each selected recording is then analyzed using three randomly selected, non-overlapping signal segments of equal length. For an arbitrary signal segment \(x[n]\) containing \(N\) complex I/Q samples, the average linear power is defined as
$$
P=\frac{1}{N}\sum_{n=0}^{N-1}|x[n]|^2.
$$
The corresponding power in dBFS is $P^{\mathrm{dBFS}}=10\log_{10}(P)$. 

Let $P_{\mathrm{env}}$ denote the environmental noise power. For a UAV-containing recording, the received signal power \(P_{\mathrm{rec}}\) contains both the UAV signal component and the environmental noise. Therefore, the valid signal power is obtained by subtracting the corresponding environmental noise power:
$$
P_{\mathrm{sig}}=\max\left(P_{\mathrm{rec}}-P_{\mathrm{env}},\epsilon\right),
$$
where \(\epsilon\) is a small positive constant. The SNR is computed as
$$
\mathrm{SNR}^{\mathrm{dB}}
=
10\log_{10}
\left(
\frac{P_{\mathrm{sig}}}{P_{\mathrm{env}}}
\right).
$$

Then, the segment is further divided into 1-ms frames. For the $m$-th frame, the average power is defined as

$$
P_{m}=\frac{1}{L}\sum_{n=mL}^{(m+1)L-1}|x[n]|^2,
$$
where \(L=F_s\times1\,\mathrm{ms}\), while \(F_s\) denotes the sampling rate. The within-segment power variability is measured using both the standard deviation and the IQR of the frame-level power:
\begin{align*}
\sigma_{P}&=\mathrm{std}\left(P_{m}^{\mathrm{dBFS}}\right),\\
\mathrm{IQR}_{P}
&=
Q_{75}\left(P_{m}^{\mathrm{dBFS}}\right)
-
Q_{25}\left(P_{m}^{\mathrm{dBFS}}\right),
\end{align*}
where \(Q_{75}(\cdot)\) and \(Q_{25}(\cdot)\) denote the 75th and 25th percentiles, respectively. 

The observable bandwidth is computed from the Welch power spectral density (PSD) estimate. To avoid overestimating the occupied bandwidth due to broadband environmental noise, the environmental PSD is subtracted from the received-signal PSD:
$$
S_{\mathrm{sig}}(f)
=
\max
\left(
S_{\mathrm{rec}}(f)-S_{\mathrm{env}}(f),0
\right).
$$
The 99\% observable bandwidth is then defined as the frequency interval containing 99\% of the cumulative signal PSD energy:
$$
B_{99}=f_{\mathrm{high}}-f_{\mathrm{low}},
$$
where
$$
\frac{
	\int_{-\infty}^{f_{\mathrm{low}}}S_{\mathrm{sig}}(f)\,df
}{
	\int_{-\infty}^{\infty}S_{\mathrm{sig}}(f)\,df
}
=0.005,
\qquad
\frac{
	\int_{-\infty}^{f_{\mathrm{high}}}S_{\mathrm{sig}}(f)\,df
}{
	\int_{-\infty}^{\infty}S_{\mathrm{sig}}(f)\,df
}
=0.995.
$$
Accordingly, $B_{99}$ represents the observable occupied bandwidth in the passively recorded RF data.

Finally, let $j=1,\ldots,J$ denote the non-overlapping segments randomly selected for each recording. For a generic metric $z \in \{P_{\mathrm{env}}, P_{\mathrm{sig}}, \mathrm{SNR}, \sigma_P, \mathrm{IQR}_P, B_{99}\}$, let $z_{j}$ denote the value obtained from the $j$-th segment. The within-record standard deviation for each recording is computed as:
$$
\sigma_{z} = \sqrt{ \frac{1}{J-1} \sum_{j=1}^{J} \left( z_{j} - \mu_{z} \right)^2 }, \quad \mu_{z}=\frac{1}{J} \sum_{j=1}^{J} {z}_{j}
$$

The quantitative results are summarized in Table~5. In the outdoor scenario, the effective received-signal power is \(-34.57 \pm 1.24\)~dBFS, compared with an environmental background level of \(-60.09 \pm 0.80\)~dBFS, yielding an SNR of \(25.47 \pm 1.24\)~dB. For the indoor scenario, the corresponding values are \(-15.39 \pm 0.19\)~dBFS, \(-36.26 \pm 0.20\)~dBFS, and \(20.87 \pm 0.19\)~dB, respectively. In both representative recordings, the UAV-related signal remains well above the corresponding environmental background, indicating clearly distinguishable RF observations. The higher valid signal and background signal power observed indoors is also consistent with the denser RF activity in the mixed WiFi-UAV environment.

The 99\% observable bandwidth is \(25.00 \pm 0.07\)~MHz outdoors and \(28.60 \pm 10.14\)~MHz indoors. Given that \(B_{99}\) is derived from the background-subtracted PSD, it represents the effective spectral occupancy observed in the passive recordings rather than the nominal transmitter or acquisition bandwidth. The nearly unchanged outdoor bandwidth across the three selected segments indicates stable spectral occupancy, whereas the substantially larger variation indoors suggests stronger segment-dependent spectral activity in the mixed WiFi-UAV environment.

The frame-level power statistics further characterize short-term temporal variation. The outdoor recording yields \(\sigma_P=7.20 \pm 0.40\)~dB and \(\mathrm{IQR}_P=1.94 \pm 0.61\)~dB, while the indoor recording yields \(5.75 \pm 0.76\)~dB and \(6.80 \pm 1.84\)~dB, respectively. The larger outdoor \(\sigma_P\) indicates stronger overall frame-level power dispersion, whereas the larger indoor IQR indicates a broader central power distribution. The within-record standard deviations quantify the sensitivity of each metric to the selection of different non-overlapping segments from the same recording. Relatively small variations are observed for signal power, SNR, and several power-statistical metrics, while the indoor observable bandwidth exhibits larger segment dependence. In general, their moderate magnitudes indicate that the extracted measurements are reasonably repeatable within individual recordings, while preserving the distinct signal characteristics of the outdoor and indoor RF environments.

\begin{table}[t]
	\centering
	\caption{Technical validation metrics for representative outdoor and indoor recordings.}
	\label{tab:indoor_outdoor_metrics}
	\begin{tabular}{ccc}
		\hline
		\textbf{Metric} & \textbf{Outdoor} & \textbf{Indoor} \\
		\hline
		$\mu_{P_{\mathrm{env}}} \pm \sigma_{P_{\mathrm{env}}}$ / dBFS
		& $-60.09 \pm 0.80$
		& $-36.26 \pm 0.20$ \\
		
		$\mu_{P_{\mathrm{sig}}} \pm \sigma_{P_{\mathrm{sig}}}$ / dBFS
		& $-34.57 \pm 1.24$
		& $-15.39 \pm 0.19$ \\
		
		$\mu_{\mathrm{SNR}} \pm \sigma_{\mathrm{SNR}}$ / dB
		& $25.47 \pm 1.24$
		& $20.87 \pm 0.19$ \\
		
		$\mu_{\sigma_P} \pm \sigma_{\sigma_P}$ / dB
		& $7.20 \pm 0.40$
		& $5.75 \pm 0.76$ \\
		
		$\mu_{\mathrm{IQR}_P} \pm \sigma_{\mathrm{IQR}_P}$ / dB
		& $1.94 \pm 0.61$
		& $6.80 \pm 1.84$ \\
		
		$\mu_{B_{99}} \pm \sigma_{B_{99}}$ / MHz
		& $25.00 \pm 0.07$
		& $28.60 \pm 10.14$ \\
		
		\hline
	\end{tabular}
	
\end{table}

\subsection*{Flight State Awareness Experiment}
To further evaluate the quality of Dataset-1 for RFF-related research, we conduct a 26-class UAV flight-state classification experiment using an EfficientNet-B0 model. The detailed experimental configuration is summarized in Table~6.
\begin{table}
		\centering
		\caption{Experimental configuration for the flight-state awareness experiment.}
		\label{tab:flight_state_config}
		\begin{tabular}{cc}
			\hline
			\textbf{Item} & \textbf{Setting} \\
			\hline
			STFT window & Hamming window \\
			Window length & 448 samples \\
			Overlap ratio & 50\% \\
			Input representation & $224 \times 224$ STFT spectrogram \\
			Backbone model & EfficientNet-B0 \\
			Optimizer & AdamW \\
			Learning rate & $3 \times 10^{-4}$ to $10^{-6}$ with cosine annealing \\
			Training epochs & 20 \\
			Batch size & 32 for training and 128 for validation and testing \\
			Evaluation metrics & Top-1 Accuracy, Top-5 Accuracy, and Macro-F1 \\
			Reproducibility & Fixed random seed \\
			\hline
		\end{tabular}
\end{table}

Two comparative experiments are designed: a baseline experiment and a background-noise calibration experiment that uses Dataset-7 as the ambient noise reference. The two experiments differ only in the data preprocessing stage. In the baseline experiment, the raw IQ recordings are directly segmented into $1~\mathrm{ms}$ clips, transformed into STFT spectrograms, and resized to $224 \times 224$. In the background-noise calibration experiment, the power statistics of the asynchronous outdoor background (10th, 25th, 50th, 75th, and 90th percentiles, and the median absolute deviation) are first used to construct three complementary STFT-domain views: background-aware spectral shrinkage, contrast-preserving spectral soft-thresholding, and conservative high-band guided enhancement. All gains are applied in the complex STFT domain with the phase preserved, after which the same segmentation, STFT transformation, and resizing procedures as those in the baseline experiment are applied. For each view, an EfficientNet-B0 is trained using the same optimizer, learning-rate schedule, batch size, and data splits as the baseline, and the final prediction combines the three branches with the fixed baseline using convex weights selected based on the validation set.
\begin{figure}
	\centering
	\includegraphics[width=16cm]{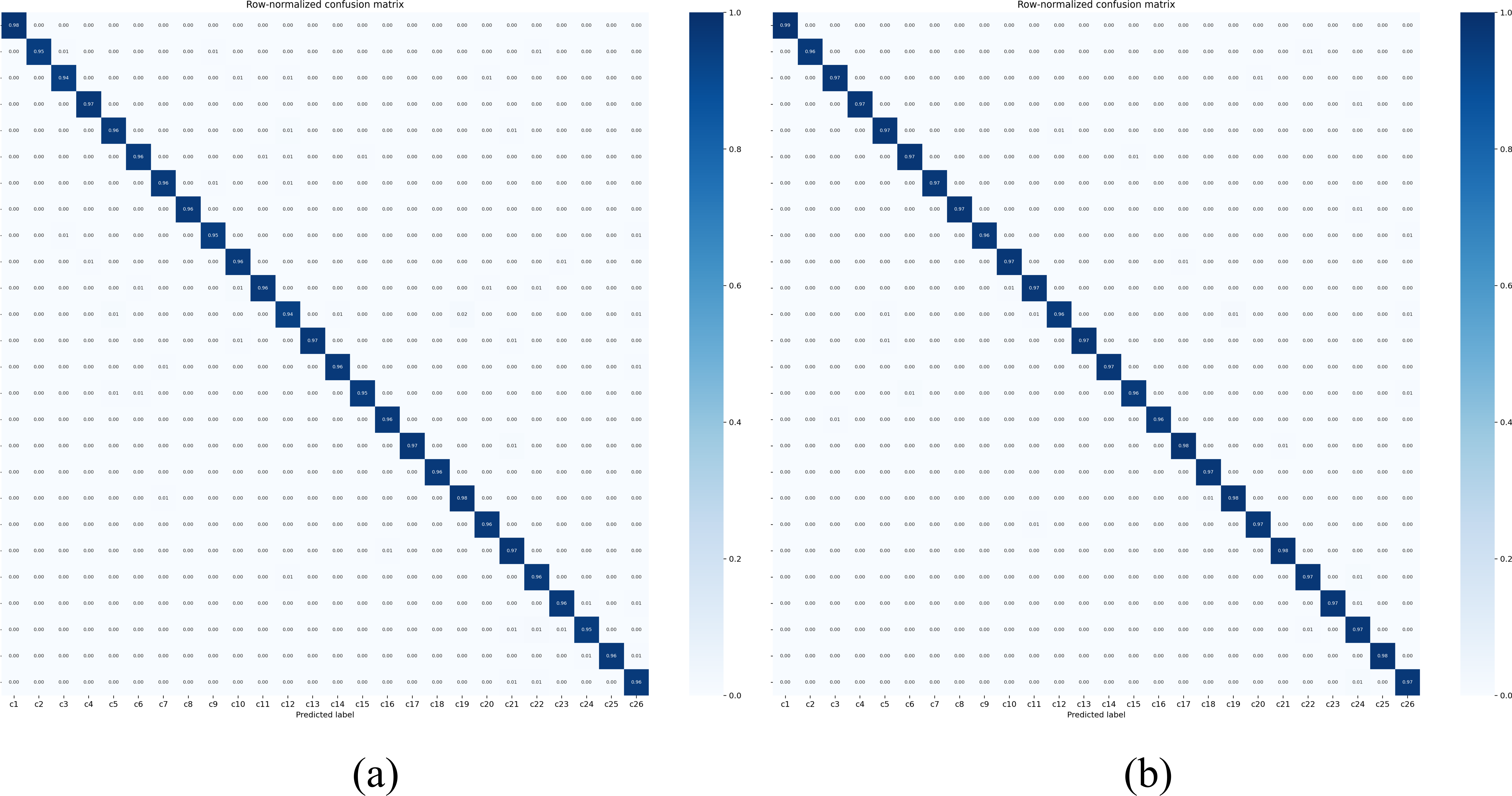}
	\caption{Row-normalized confusion matrices for the 26-class UAV flight-state classification results.
		(a) EfficientNet-B0 baseline experiment, with a Top-1 Accuracy of 96.04\%.
		(b) Background-noise calibration experiment, with a Top-1 Accuracy of 97.11\%.}
	\label{fig:classification}
\end{figure}
For both experiments, all data are divided into training, validation, and testing sets with a ratio of 6:2:2. The final results show that the baseline experiment achieves 96.04\% Top-1 Accuracy, 98.48\% Top-5 Accuracy, and a Macro-F1 score of 96.05\%. The background-noise calibration experiment achieves 97.11\% Top-1 Accuracy, 98.88\% Top-5 Accuracy, and a Macro-F1 score of 97.11\%.
The corresponding row-normalized confusion matrices are shown in Figure~4. These results indicate that background noise still provides obvious influence for the classification task.

\section*{Dataset Limitations}

\subsection*{Lack of Physical-Layer Metadata and Relation to 3GPP ISAC-UAV Scenarios}

It is worth noting that the current release of DRFF-R2 does not include detailed physical-layer metadata, such as waveform and modulation configurations, or standardized channel parameters. This limitation mainly originates from the closed-source nature of commercial UAV platforms, for which the internal communication configurations and protocol-specific physical-layer procedures are generally inaccessible to external authorities. Therefore, the dataset does not provide the protocol-level information or standardized reference measurements required to directly reproduce 3GPP-compliant procedures such as, reference-signal-based channel estimation, positioning, or link-level communication evaluation.

This limitation is consistent with the intended application of DRFF-R2. The dataset is designed for passive UAV monitoring scenarios, where consumer UAVs are usually treated as unknown RF emitters and the monitoring platform typically has no access to prior knowledge about their internal communication protocols, signaling structures, or physical-layer configurations. Accordingly, DRFF-R2 focuses on passive RF monitoring tasks, including real-time UAV identification, operational-state recognition, and blind signal separation from externally observable signal characteristics, rather than on protocol-specific communication performance evaluation.

Despite this distinction, the collection design of DRFF-R2 covers  several measurement dimensions that are closely related to UAV wireless research scenarios considered in 3GPP and ISAC studies. For example, 3GPP TR 36.777 investigates aerial-user communication under altitude-dependent propagation and aerial interference conditions, while the ISAC-UAV scenarios introduced in 3GPP TR 38.901 consider outdoor UAV sensing targets under LoS/NLoS propagation, three-dimensional mobility, different UAV heights, and multiple sensing targets. Correspondingly, DRFF-R2 includes measurements under different UAV altitudes and flight velocities, LoS/NLoS propagation conditions,  Doppler-related temporal variations, multi-UAV co-channel interference, and realistic electromagnetic background environments. These characteristics provide practical RF observations for data-driven UAV sensing, aerial interference analysis, and low-altitude ISAC.

\subsection*{Limited Hardware Calibration and Signal-Level Preprocessing}
Another limitation of the current DRFF-R2 release is the lack of dedicated hardware calibration, synchronization calibration, and background-channel compensation during data acquisition. Since DRFF-R2 was collected using a passive RF monitoring configuration, the transmitter-side hardware parameters, baseband reference signals, and calibration sequences of the monitored commercial UAVs were unavailable. Under such conditions, RF fingerprints, channel-dependent features, receiver-side effects, and synchronization-related errors cannot be completely isolated from each other. Therefore, DRFF-R2 provides realistic passive RF observations that capture both transmitter-dependent characteristics and environmental effects, rather than calibrated transmitter-only RF fingerprints. This design is consistent with practical UAV monitoring scenarios, where consumer UAVs are usually observed as unknown RF emitters and the monitoring platform has no access to their internal communication or calibration information.

\subsection*{Limited Recording-Level Independence}
A further limitation of the current dataset is that, for a given UAV instance and operating condition, the available samples are generally derived from a single continuous RF recording. Although the training, validation, and test subsets can be constructed using non-overlapping signal segments, samples originating from the same recording cannot be regarded as strictly statistically independent. They may share slowly varying recording-level characteristics, such as propagation conditions, interference background, and hardware state. Nevertheless, the lack of strict recording-level independence does not imply that different segments from the same recording are identical or devoid of useful variability. Non-overlapping temporal segments may still exhibit meaningful variations in received power, spectral characteristics, and interference conditions, as also supported by the results presented in the Signal Quantitative Characterization and Validation section. In addition, Datasets-1 and Datasets-3 contain cross-day recordings of the same UAV devices acquired in separate recording sessions. These independently acquired recordings provide a complementary basis for investigating time-varying signal properties across acquisition sessions while preserving the underlying device-related RF fingerprint information. Therefore, the current dataset remains useful for studying within-record discriminability, temporal signal variability, and robustness to condition changes.

\section*{Data Availability}
The dataset is publicly hosted on the Science Data Bank platform at \url{https://doi.org/10.57760/sciencedb.36815}~[25], ensuring long-term preservation, accessibility, and reproducibility. This dataset is distributed under the CC BY 4.0 license.

\section*{Code Availability}
The signal acquisition program and associated experimental code are publicly available at \url{https://doi.org/10.57760/sciencedb.36815}

\section*{Author Contribution}
H.Z., N.G., Z.Z., and Z.H. jointly contributed to the construction and testing of the overall dataset. H.Z. led the investigation, curated and managed the dataset, and drafted the main manuscript. N.G. contributed to writing—review and editing, funding acquisition, and project administration. Z.Z. and Z.H. jointly conducted the experiments and analysis for the Technical Validation section. S.J. contributed to supervision, conceptualization, and writing—review and editing. M.M. contributed to writing—review and editing, and formal analysis.

\section*{Acknowledgements}
This work was supported in part by the National Key Research and
Development Program of China under Grant 2024YFE0200700, in part by
National Science Foundation of China (NSFC) under Grants 62371131, and
in part by Undergraduate Training Programs for Innovation of  Jiangsu Province.


\begin{thebibliography}{99}
	
	\bibitem{ref1}
	He, Y., Zhang, J., Xi, R., Na, X., Sun, Y. \& Li, B. Detection and identification of non-cooperative UAV using a COTS mmWave radar. \emph{ACM Trans. Sens. Netw.} \textbf{20}, 44 (2024). https://doi.org/10.1145/3638767
	
	\bibitem{ref2}
	Yang, S., Han, H., Liu, Y., Guo, W., Pang, Z. \& Zhang, L. Reconfigurable intelligent surface-induced randomness for mmWave key generation. In \emph{Proc. IEEE ICC} 2909–2914 (2023). https://doi.org/10.1109/ICC45041.2023.10278950
	
	\bibitem{ref3}
	Yu, N., Mao, S., Zhou, C., Sun, G., Shi, Z. \& Chen, J. DroneRFa: A large-scale dataset of drone radio frequency signals for detecting low-altitude drones. \emph{J. Electron. Inf. Technol.} \textbf{46}, 1147–1156 (2024). https://doi.org/10.11999/JEIT230570
	
	\bibitem{ref4}
	Ren, J., Yu, N., Zhou, C., Shi, Z. \& Chen, J. DroneRFb-DIR: An RF signal dataset for non-cooperative drone individual identification. \emph{J. Electron. Inf. Technol.} (2025). https://doi.org/10.11999/JEIT240804
	
	\bibitem{ref5}
	Zheng, H., Gao, N., Cai, D., Jin, S. \& Matthaiou, M. UAV individual identification via distilled RF fingerprints-based LLM in ISAC networks. \emph{IEEE Wireless Commun. Lett.} \textbf{14}, 3769–3773 (2025). https://doi.org/10.1109/LWC.2025.3603423
	
	\bibitem{ref6}
	Shi, R. \emph{et al.} RFUAV: A benchmark dataset for unmanned aerial vehicle detection and identification. Preprint at https://arxiv.org/abs/2503.09033 (2025).
	
	\bibitem{ref7}
	Vuorenmaa, M., Marin, J., Heino, M., Turunen, M. \& Riihonen, T. Radio-frequency control and video signal recordings of drones. Zenodo https://doi.org/10.5281/zenodo.4264467 (2020).
	
	\bibitem{ref8}
	Al-Sa’d, M. F., Al-Ali, A., Mohamed, A., Khattab, T. \& Erbad, A. RF-based drone detection and identification using deep learning approaches: An initiative towards a large open source drone database. \emph{Future Gener. Comput. Syst.} \textbf{100}, 86–97 (2019). https://doi.org/10.1016/j.future.2019.05.007
	
	\bibitem{ref9}
	Irfan, M., Dalai, S., Vishwakarma, K., Trslic, P., Riordan, J. \& Dooly, G. Multi-sensor fusion for efficient and robust UAV state estimation. In \emph{Proc. ICCMA} 35–40 (2024). https://doi.org/10.1109/ICCMA63715.2024.10843888
	
	\bibitem{ref10}
	Yang, S., Luo, Y., Miao, W., Ge, C., Sun, W. \& Luo, C. RF signal-based UAV detection and mode classification: A joint feature engineering generator and multi-channel deep neural network approach. \emph{Entropy} \textbf{23}, 1678 (2021). https://doi.org/10.3390/e23121678
	
	\bibitem{ref11}
	Zuo, M., Xie, S., Zhang, X. \& Yang, M. Recognition of UAV video signal using RF fingerprints in the presence of WiFi interference. \emph{IEEE Access} \textbf{9}, 88844–88851 (2021). https://doi.org/10.1109/ACCESS.2021.3089590
	
	\bibitem{ref12}
	Ezuma, M., Erden, F., Anjinappa, C. K., Ozdemir, O. \& Guvenc, I. Detection and classification of UAVs using RF fingerprints in the presence of Wi-Fi and Bluetooth interference. \emph{IEEE Open J. Commun. Soc.} \textbf{1}, 60–76 (2020). https://doi.org/10.1109/OJCOMS.2019.2955889
	
	\bibitem{ref13}
	Medaiyese, O. O., Ezuma, M., Lauf, A. P. \& Adeniran, A. A. Hierarchical learning framework for UAV detection and identification. \emph{IEEE J. Radio Freq. Identif.} \textbf{6}, 176–188 (2022). https://doi.org/10.1109/JRFID.2022.3157653
	
	\bibitem{ref14}
	Medaiyese, O. O., Ezuma, M., Lauf, A. P. \& Guvenc, I. Wavelet transform analytics for RF-based UAV detection and identification system using machine learning. \emph{Pervasive Mob. Comput.} \textbf{82}, 101569 (2022). https://doi.org/10.1016/j.pmcj.2022.101569
	
	\bibitem{ref15}
	Alhazbi, S., Sciancalepore, S. \& Oligeri, G. The day-after-tomorrow: On the performance of radio fingerprinting over time. In \emph{Proc. ACSAC} 439–450 (2023). https://doi.org/10.1145/3627106.3627192
	
	\bibitem{ref16}
	He, J., Huang, S., Chang, S., Wang, F., Shen, B.-Z. \& Feng, Z. Radio frequency fingerprint identification with hybrid time-varying distortions. \emph{IEEE Trans. Wirel. Commun.} \textbf{22}, 6724–6736 (2023). https://doi.org/10.1109/TWC.2023.3245070
	
	\bibitem{ref17}
	Gao, N., Liu, Y., Zhang, Q., Li, X. \& Jin, S. Let RFF do the talking: large language model enabled lightweight RFFI for 6G edge intelligence. \textit{Sci. China Inf. Sci.} \textbf{68}, 170308 (2025). https://doi.org/10.1007/s11432-024-4463-0
	
	\bibitem{ref18}
	Liu, Y. et al. Self-robust RF fingerprint identification with 6G agentic AI: cloud-edge intelligent collaboration. \textit{IEEE Trans. Cogn. Commun. Netw.} \textbf{12}, 8718--8732 (2026). https://doi.org/10.1109/TCCN.2026.3703008
	
	\bibitem{ref19}
	Kim, M., Choi, W., Chung, J., Lee, D. \& Jung, S. KUIELab-MDX-Net: a two-stream neural network for music demixing. Preprint at https://arxiv.org/abs/2111.12203 (2021).
	
	\bibitem{ref20}
	Rouard, S., Massa, F. \& D'efossez, A. Hybrid transformers for music source separation. In \textit{Proc. IEEE Int. Conf. Acoustics, Speech and Signal Processing (ICASSP)} 1--5 (2023). https://doi.org/10.1109/ICASSP49357.2023.10096956
	
	\bibitem{ref21}
	Stoller, D., Ewert, S. \& Dixon, S. Wave-U-Net: a multi-scale neural network for end-to-end audio source separation. Preprint at https://arxiv.org/abs/1806.03185 (2018)
	
	\bibitem{ref22}
	Chen, Z., Pang, Z., Hou, W., Wen, H., Wen, M., Zhao, R. \& Tang, T. Cross-device radio frequency fingerprinting identification based on domain adaptation. \textit{IEEE Trans. Consum. Electron.} \textbf{70}, 2391--2400 (2024). https://doi.org/10.1109/TCE.2024.3357844
	
	\bibitem{ref23}
	Bousmalis, K., Trigeorgis, G., Silberman, N., Krishnan, D. \& Erhan, D. Domain separation networks. \textit{Adv. Neural Inf. Process. Syst.} \textbf{29}, 343--351 (2016). https://arxiv.org/abs/1608.06019
	
	\bibitem{ref24}
	Wang, S., Wan, H., Wang, F., Wang, Y., Lin, Y. \& Gui, G. TFMix: a robust time-frequency mixing approach for domain generalization in specific emitter identification. \textit{IEEE Trans. Cogn. Commun. Netw.} \textbf{11}, 3981--3992 (2025). https://doi.org/10.1109/TCCN.2025.3541723
	
	\bibitem{ref25}
	Zheng, H. \& Gao, N. DRFF-R2: Multi-Scenario UAV Radio Frequency Signal Dataset.
	Science Data Bank (2026). https://doi.org/10.57760/sciencedb.36815.
	
	\bibitem{ref26}
	Yucek, T. \& Arslan, H. A survey of spectrum sensing algorithms for cognitive radio applications. 
	\textit{IEEE Commun. Surv. Tutor.} \textbf{11}, 116--130 (2009).
	https://doi.org/10.1109/SURV.2009.090109
	
\end{thebibliography}
\end{document}